\documentclass[conference]{IEEEtran}
\IEEEoverridecommandlockouts
\usepackage{makecell}
\usepackage{cite}
\usepackage{amsmath,amssymb,amsfonts}
\usepackage{float}  
\usepackage{placeins}
\usepackage{graphicx}
\usepackage{textcomp}
\usepackage{xcolor}
\usepackage{algorithm}
\usepackage{algorithmic}
\usepackage{enumitem}

\usepackage{preambles}

\makeatletter
\newcommand{\linebreakand}{%
  \end{@IEEEauthorhalign}
  \hfill\mbox{}\par
  \mbox{}\hfill\begin{@IEEEauthorhalign}
}
\makeatother

\usepackage[caption=false,font=footnotesize]{subfig}
\def\BibTeX{{\rm B\kern-.05em{\sc i\kern-.025em b}\kern-.08em
    T\kern-.1667em\lower.7ex\hbox{E}\kern-.125emX}}
\begin{document}

\title{Aligning Microscopic Vehicle and Macroscopic Traffic Statistics: Reconstructing Driving Behavior from Partial Data\\
{}
\thanks{Zhihao Zhang and Keith A. Redmill are supported by  Carnegie Mellon University’s Safety21 National University Transportation Center, which is sponsored by the US Department of Transportation under grants 69A3552344811/69A3552348316.}
}

\author{
\IEEEauthorblockN{Zhihao Zhang}
\IEEEauthorblockA{\textit{dept. Electrical and Computer Engineering}\\
\textit{The Ohio State University}\\
Columbus, USA\\
zhang.11606@osu.edu}
\and
\IEEEauthorblockN{Keith A. Redmill}
\IEEEauthorblockA{\textit{dept. Electrical and Computer Engineering}\\
\textit{The Ohio State University}\\
Columbus, USA\\
redmill.1@osu.edu}
\linebreakand
\IEEEauthorblockN{Chengyang Peng}
\IEEEauthorblockA{\textit{dept. Mechanical and Aerospace Engineering}\\
\textit{The Ohio State University}\\
Columbus, USA\\
peng.947@osu.edu}
\and
\IEEEauthorblockN{Bowen Weng}
\IEEEauthorblockA{\textit{Department of Computer Science}\\
\textit{Iowa State University}\\
Ames, IA, USA\\
bweng@iastate.edu}
}
\maketitle

\begin{abstract}
A driving algorithm that aligns with good human driving practices, or at the very least collaborates effectively with human drivers, is crucial for developing safe and efficient autonomous vehicles. In practice, two main approaches are commonly adopted: (i) supervised or imitation learning, which requires comprehensive naturalistic driving data capturing all states that influence a vehicle's decisions and corresponding actions, and (ii) reinforcement learning (RL), where the simulated driving environment either matches or is intentionally more challenging than real-world conditions. Both methods depend on high-quality observations of real-world driving behavior, which are often difficult and costly to obtain.
State-of-the-art sensors on individual vehicles can gather microscopic data, but they lack context about the surrounding conditions. Conversely, roadside sensors can capture traffic flow and other macroscopic characteristics, but they cannot associate this information with individual vehicles on a microscopic level. 
Motivated by this complementarity, we propose a framework that reconstructs unobserved microscopic states from macroscopic observations, using microscopic data to anchor observed vehicle behaviors, and learns a shared policy whose behavior is microscopically consistent with the partially observed trajectories and actions and macroscopically aligned with target traffic statistics when deployed population-wide. Such  constrained and regularized policies promote realistic flow patterns and safe coordination with human drivers at scale. 


\end{abstract}


\section{Introduction}
\begin{figure}[t]
\centering
\includegraphics[width=\columnwidth]{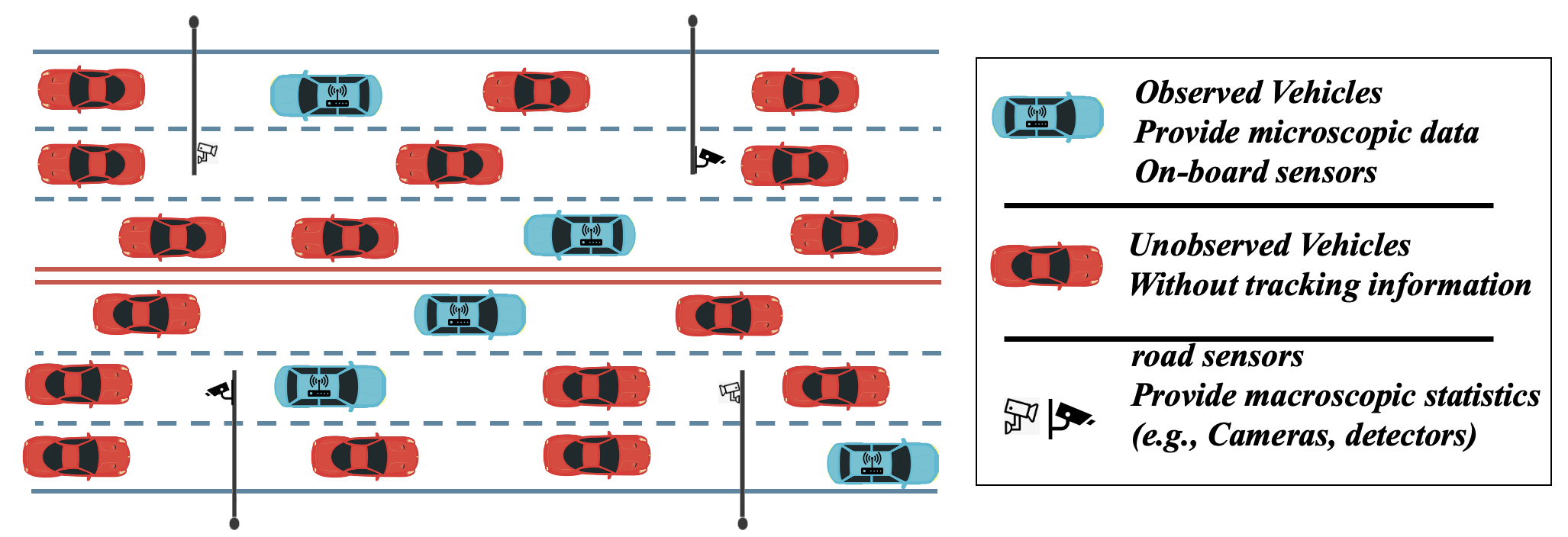}
\caption{Partially observed traffic setting. A small subset of instrumented vehicles (blue) provides microscopic trajectories and actions, while the remaining vehicles (red) are unobserved. Roadside sensors (e.g., cameras/detectors) supply macroscopic statistics for the whole stream}

\label{fig:statement}
\end{figure}

The ability to design driving algorithms that emulate human behavior and coordinate safely with human drivers depends in part on how driving data are collected and used. 
In practice, such data are organized at two complementary levels as shown in Fig.~\ref{fig:statement}. Microscopic data capture individually observed vehicle trajectories, control actions, and local surroundings using on-board sensors~\cite{chen2015deepdriving,bojarski2016end,xu2017end,hu2023planning}. These data are the basis for supervised and imitation learning pipelines that fit a parameterized policy to map local observations to human-like actions~\cite{han2019driving,bojarski2016end,bojarski2017explaining,codevilla2018end,hawke2020urban,chen2018deep,sonu2018exploiting}. These approaches can produce highly interpretable and reactive policies. However, they critically rely on dense, high-resolution sensing that observes all relevant agents and typically require substantial, labor-intensive labeling~\cite{bojarski2016end,geiger2012are,paden2016survey}. Macroscopic data, on the other hand, consist of aggregated traffic statistics such as mean speed, flow, and density over space and time, typically obtained from infrastructure sensors such as cameras or loop detectors~\cite{helbing2001traffic}. Such aggregates are widely available and are routinely used to calibrate or validate large-scale traffic simulators and reinforcement learning environments~\cite{kendall2019learning,codevilla2018end}, where background vehicles follow dynamics tuned to match observed flow patterns. While RL in these simulators can optimize long-term performance under challenging conditions, its fidelity is limited by how well the simulated population reproduces real macroscopic behavior and by the lack of direct association with individual human trajectories~\cite{yurtsever2020survey,kiran2021deep,treiber2000congested}.

Methods built exclusively on microscopic trajectories from instrumented or connected vehicles can fit fine-grained state–action maps but suffer from limited coverage and missing population context, as such traces represent only a small fraction of the flow~\cite{traffictechnologytoday2019,itskrs2022,traffictechnologytoday2018,usdot2023,verizon2024}. In contrast, methods calibrated solely on infrastructure aggregates (e.g., mean speed, flow, density) reproduce bulk trends yet cannot attribute behavior to individual agents, creating an identifiability gap in which multiple policies yield indistinguishable aggregates~\cite{djuric2018short,chen2019deep}.

Motivated by this complementarity, we propose a macro–micro integrated framework that reconstructs unobserved microscopic states from macroscopic observations and learns a shared policy whose rollouts are jointly consistent with individual demonstrations and aggregate measurements. Such frameworks should use available microscopic data to anchor observed individual behavior.
Macroscopic statistics can then be used to constrain and regularize policies, promoting realistic flow patterns and safe coordination with human drivers at scale. Prior works on hybrid imitation learning and RL have begun to explore this direction~\cite{kiran2021deep,zhang2021robustness}, but they typically assume either richly instrumented agents or idealized simulators and do not fully address learning from heterogeneous, partially observed real-world data streams.
\begin{figure}[!t]
\centering
\includegraphics[width=\columnwidth]{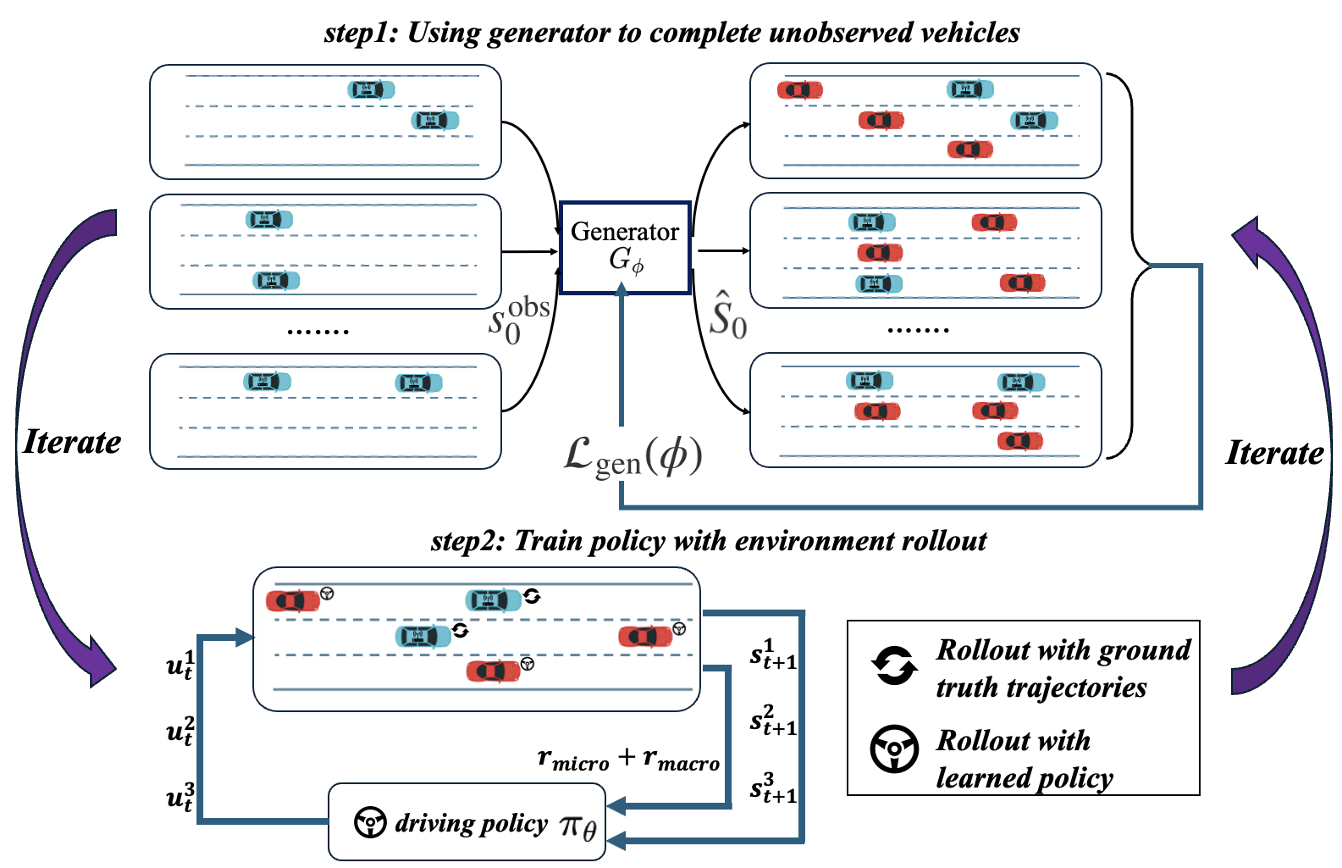}
\caption{Two-stage framework. Top (Step~1): the generator \(G_\phi\) takes partial snapshots \(s_0^{\mathrm{obs}}\), completes hidden vehicles to form \(\hat{S}_0\), and is trained with the generator loss \(\mathcal{L}_{\mathrm{gen}}(\phi)\). Bottom (Step~2): episodes start from \(\hat{S}_0\); the shared policy \(\pi_\theta\) rolls out the environment and is updated using a trajectory-level score combining microscopic imitation \(r_{\mathrm{micro}}\) and macroscopic alignment \(r_{\mathrm{macro}}\). In each simulation, observed vehicles follow their ground-truth trajectories, while unobserved vehicles are controlled by the learned policy.}
\vspace{-6mm}
\label{fig:solution2}
\end{figure}

\subsection{Main Contributions}
\noindent The main contributions of this work can be summarized into the following three aspects:
\begin{itemize}[leftmargin=*]
    \item \textbf{Problem formulation:} We aim to learn one shared driving policy using partial micro data and macro traffic summaries. The resulting policy should fit observed actions and reproduce aggregate statistics.

    \item \textbf{Two-stage solution:} We propose a two-stage policy training framework:
    (i) a generator that uses partial microscopic data and macroscopic statistics to  generate unobserved vehicles,
    and (ii) a RL policy learning scheme that trains a shared policy on these completed scenes
    so that its behavior matches individual demonstrations and reproduces the observed traffic statistics.

    \item\textbf{Case study:} We implement this framework in a ring-road case study with concrete generator and policy architectures, and show that the learned policy matches target speed and spacing statistics while producing realistic car-following trajectories at the microscopic level.
\end{itemize}

\section{Preliminaries and Problem Formulation}\label{sec:prob}

Before defining the shared driving policy, we introduce the basic traffic notation.
We model traffic as a multi-agent dynamical system with a number \(N\) of vehicles indexed by \(\mathcal{I}=\{1,\dots,N\}\).
At time \(t\), vehicle \(i\) has state \(s_t^i \in \mathbb{R}^{d_s}\) and applies control \(u_t^i \in \mathbb{R}^{d_u}\).
The dynamics are \(s_{t+1}^i = f(s_t^i, u_t^i; \omega)\), where \(f\) is the state update model and \(\omega\) denotes model uncertainty.
The joint microscopic state is \(S_t=\{s_t^i\}_{i=1}^N\).
Let the time index set be \(\mathcal{T}=\{0,1,\dots,T-1\}\).
When clear from context, we omit the time subscript and write \(S\) for a generic state at \(t \in \mathcal{T}\).

From the individual-vehicle perspective, a driving policy must depend on locally observable information.
Let \(O(S,i)\in\mathbb{R}^{d_o}\) be vehicle \(i\)’s observation (e.g., ego state and nearby vehicles).
Each vehicle selects an action from its observation under a stochastic driving policy \(\pi_{\theta}: S \mapsto u\).
Given the common acknowledgment that behaviors are statistically aligned across vehicles~\cite{helbing2002micro,kesting2008calibrating,treiber2025car}, we posit a shared policy \(\pi_\theta\) such that the action of any vehicle is \(u^i=\pi_\theta(O(S,i))\) for all \(i\in\mathcal{I}\).
\enlargethispage{-\baselineskip}
With the shared policy in place, we now describe the two complementary partial-data
views used in this work: a microscopic view based on instrumented vehicles and a
macroscopic view based on infrastructure sensing.

For the microscopic view, only a small subset of vehicles carries on-board sensors.
We denote the index set of these instrumented vehicles by
$\mathcal{I}_{\mathrm{obs}}\subset\mathcal{I}$, where
$|\mathcal{I}_{\mathrm{obs}}|\ll N$ indicates that this subset is a small fraction
of the $N$ vehicles on the road. In other words, $\mathcal{I}_{\mathrm{obs}}$ collects
the IDs of vehicles for which both observations and actions are available.
The microscopic dataset is
$\mathcal{D}_{\mathrm{micro}}=\{(O(S,i),\,u^i)\mid i\in\mathcal{I}_{\mathrm{obs}}\}$.
Standard imitation learning fits a parameterized policy $\pi_\theta$ using only
$\mathcal{D}_{\mathrm{micro}}$, i.e.,
$\pi_\theta\in\arg\min_{\theta}\sum_{(o,u)\in\mathcal{D}_{\mathrm{micro}}}
d\big(\pi_\theta(o),\,u\big)$, where $d$ denotes a pointwise discrepancy.

For the macroscopic view, infrastructure sensors provide scene-level aggregate measures of traffic behavior.
We use a feature map $\Psi:\mathcal{S}\rightarrow\mathbb{R}^{p}$ and write $\Psi(S)$
for descriptors such as mean speed, density, flow, and headway statistics over the
road segment. These quantities are commonly obtained from loop detectors, radar,
and cameras deployed on major corridors.

Together, these two sets of observationsare complementary. \(\mathcal{D}_{\mathrm{micro}}\) supplies action‐level supervision on an instrumented subset, while \(\Psi(S)\) summarizes the full traffic stream.

Our goal is to learn a shared policy $\pi_\theta$ that is microscopically consistent with observed agent behavior and macroscopically aligned with aggregate traffic statistics. Microscopic and macroscopic consistency as defined as follow:
\begin{enumerate}[leftmargin=*]
    \item Microscopically consistent on the observed vehicles means minimizing the distance between predicted and observed actions: $\min_{\theta}\sum_{i\in\mathcal{I}_{\mathrm{obs}}}
    \sum_{t\in\mathcal{T}}
    d\Big(\pi_\theta(O(S_t,i)),\, u_t^i\Big)$, where $d(\cdot,\cdot)$ is a distance between the two actions. 
    \item Macroscopic consistency means that, when \(\pi_\theta\) is followed population-wide, the induced trajectory \(\{S_t^{\pi_\theta}\}_{t \in \mathcal{T}}\) matches the target aggregate statistics: $\min_{\theta}
      \sum_{t\in\mathcal{T}}
      d
      \Big(
        \Psi(S_t^{\pi_\theta}),\,
        \Psi(S)
      \Big)$, where $d(\cdot,\cdot)$ is a distance between the two statistics.
\end{enumerate}

In our case study, both microscopic and macroscopic consistency are enforced during training: the generator is optimized with \(\mathcal{L}_{\mathrm{gen}}(\phi)\), and the policy is trained with the trajectory-level objective \(J(\theta)\).

To the best of our knowledge, jointly enforcing micro-macro alignment from heterogeneous, partially observed data has not been explicitly addressed and is challenging due to partial observability, identifiability, and multi-agent coupling.

\section{Proposed Solution}
\label{sec:proposed}
As mentioned in Section~\ref{sec:prob}, We aim to learn a shared policy \(\pi_\theta\) from partially observed microscopic trajectories and macroscopic aggregates \(\Psi(S)\) so that the learned behavior is microscopically consistent with observed actions and macroscopically aligned with target statistics when deployed population-wide.

As shown in Fig.~\ref{fig:solution2}, we realize this objective via a two-stage framework. A generator \(G_{\phi}:(s^{\mathrm{obs}},\,\Psi(S))\mapsto \hat{S}\) that reconstructs hidden microscopic states from partial observations and macroscopic statistics and a shared policy \(\pi_{\theta}:S\mapsto u\) that enforces macro--micro alignment in closed-loop rollouts. We will further discuss two stages in detail.

\subsection{Stage I: Generator for Hidden-State Completion}

\textbf{Goal.}
Given a partially observed snapshot \(s_0^{\mathrm{obs}}\) and its aggregate descriptor $\Psi(S)$ , the goal of this stage is to produce a completed initial state \(\hat{S}_0\) using a learn-based generator $G_\phi$.

\textbf{Training.}
If we define the fully observed training scenes as
\(S_0 = \{s_0^i\}_{i\in\mathcal{I}}\),
then the index set of observed vehicles is
\(\mathcal{I}_{\mathrm{obs}} \subset \mathcal{I}\),
and the index set of unobserved vehicles is
\(\mathcal{I}_{\mathrm{hid}} = \mathcal{I} \setminus \mathcal{I}_{\mathrm{obs}}\).
For each scene we know the observed vehicle states
\(s_0^{\mathrm{obs}} = \{s_0^i\}_{i\in\mathcal{I}_{\mathrm{obs}}}\),
while the unobserved vehicle states
\(s_0^{\mathrm{hid}} = \{s_0^j\}_{j\in\mathcal{I}_{\mathrm{hid}}}\)
are unknown.

To reconstruct the full scene from partial information, we use the generator to map from the known part and the macroscopic descriptor to the hidden part,
\(\hat{s}_0^{\mathrm{hid}} = G_\phi(s_0^{\mathrm{obs}}, \Psi(S))\),
and then form the completed state
\(\hat{S}_0 = \big(s_0^{\mathrm{obs}}, \hat{s}_0^{\mathrm{hid}}\big)\).
The generator is parameterized by \(\phi\) and optimized via
\begin{equation}
\mathcal{L}_{\mathrm{gen}}(\phi)
=
\lambda_{\mathrm{macro}}\,
d\big(\Psi(\hat{S}_0), \Psi(S)\big)
+
\lambda_{\mathrm{rec}}\,
d\big(\hat{s}_{t+1}^{\mathrm{hid}}, s_{t+1}^{\pi_\theta,\mathrm{hid}}\big)
\label{eq:gen-loss}
\end{equation}
where \(\lambda_{\mathrm{macro}}, \lambda_{\mathrm{rec}} \ge 0\) are weighting coefficients. Here \(s_{t+1}^{\pi_\theta,\mathrm{hid}}\) denotes the hidden state at time \(t{+}1\) obtained by rolling out the dynamics under the learned policy \(\pi_\theta\), i.e.,
\(s_{t+1}^{\pi_\theta,\mathrm{hid}} = f\big(s_t^{\mathrm{hid}},\, \pi_\theta(O(S_t));\, \omega\big)\).
The term \(\hat{s}_{t+1}^{\mathrm{hid}} = G_\phi(s_{t+1}^{\mathrm{obs}}, \Psi(S))\) is the generator’s prediction of the same hidden state using the next-step observations and macroscopic descriptor.
The reconstruction discrepancy \(d\big(\hat{s}_{t+1}^{\mathrm{hid}}, s_{t+1}^{\pi_\theta,\mathrm{hid}}\big)\) therefore encourages the generator’s completions to stay consistent with the hidden states implied by the current policy, so that completed scenes and policy rollouts remain aligned over time. This consistency term is optional: it is only activated when we iteratively refine both the generator and the policy, and can be omitted, as is done in the case study in Section~\ref{subsec:gen-arch}, if we do not perform such joint refinement. Detailed choices of the weights are reported in the case study.

\subsection{Stage II: Policy Learning with Macro--Micro Consistency}

\textbf{Goal.}
Learn a shared policy \(\pi_\theta\) that satisfies the microscopic and macroscopic
consistency objectives in Section~\ref{sec:prob}.
\enlargethispage{-\baselineskip}
\textbf{Training.}
Unlike in the traditional RL learning paradigm, we evaluate the shared policy \(\pi_\theta\) at the level of full rollouts rather than
assigning rewards state by state. To be specific,  we first initialize each episode from a
generator-completed state \(\hat{S}_0\), and then roll out the dynamics under
\(\pi_\theta\) to obtain a trajectory \(\tau = \{S_t\}_{t\in\mathcal{T}}\). On each trajectory \(\tau\), we define a microscopic score that measures how well
the policy matches the observed actions on observed vehicles:
\begin{equation}
r_{\mathrm{micro}}(\tau;\theta)
=
- \sum_{i\in\mathcal{I}_{\mathrm{obs}}}
  \sum_{t\in\mathcal{T}}
  d(\pi_\theta(O(S_t,i)),\, u_t^i).
\label{eq:r-micro}
\end{equation}
In parallel, we define a macroscopic score that measures how well the trajectory
reproduces the target aggregate statistics:
\begin{equation}
r_{\mathrm{macro}}(\tau;\theta)
=
- \sum_{t\in\mathcal{T}}
D_{\mathrm{macro}}
\Big(
\Psi(S_t^{\pi_\theta}),\,
\Psi(S)
\Big),
\label{eq:r-macro}
\end{equation}
where \(\Psi(S_t^{\pi_\theta})\) is the macroscopic feature of the simulated
state at time \(t\).
The overall policy objective combines these two trajectory-level scores is then:
\begin{equation}
J(\theta)
=
\mathbb{E}_{\tau \sim (\pi_\theta, G_\phi)}
\Big[
r_{\mathrm{micro}}(\tau;\theta)
+
\eta\, r_{\mathrm{macro}}(\tau;\theta)
\Big],
\label{eq:policy-objective}
\end{equation}
where \(\eta \ge 0\) controls the relative weight of macroscopic alignment. We optimize \(\theta\) to maximize \(J(\theta)\) using policy-gradient or
actor--critic updates over batches of rollouts, so that each update reflects
the aggregated macro--micro performance of the entire policy, rather than
treating individual transitions in isolation. The entire procedure is summarized in Algorithm~\ref{alg:macro-micro}.

\begin{algorithm}[h]
\caption{Two-stage policy learning framework}
\label{alg:macro-micro}
\begin{algorithmic}[1]
\STATE \textbf{Stage I: Train \(G_\phi\)}
\WHILE{not converged (validation generator loss $\mathcal{L}_{\mathrm{gen}}\!$ (Eqs.~\eqref{eq:gen-loss}) does not improve by $\varepsilon\!$ for sufficient number of epochs, or max steps reached)}
    \STATE Sample fully observed snapshot \(S_0\).
    \STATE Choose \(\mathcal{I}_{\mathrm{obs}}\), set \(\mathcal{I}_{\mathrm{hid}}\!=\!\mathcal{I}\setminus\mathcal{I}_{\mathrm{obs}}\).
    \STATE Form \(s_0^{\mathrm{obs}},\, s_0^{\mathrm{hid}}\); compute \(\Psi(S)\!=\!\Psi(S_0)\).
    \STATE Predict \(\hat{s}_0^{\mathrm{hid}}\!=\!G_\phi(s_0^{\mathrm{obs}},\,\Psi(S))\);
           set \(\hat{S}_0\!=\!(s_0^{\mathrm{obs}},\,\hat{s}_0^{\mathrm{hid}})\).
    \STATE Update \(\phi\) using \(\mathcal{L}_{\mathrm{gen}}(\phi)\) (Eqs.~\eqref{eq:gen-loss}).
\ENDWHILE

\STATE \textbf{Stage II: Train \(\pi_\theta\) with fixed \(G_\phi\).}
\WHILE{not converged (policy objective \(J(\theta)\!\) (Eqs.~\eqref{eq:policy-objective}) does not improve by $\varepsilon\!$ for sufficient number of steps, or max steps reached)}
    \STATE Sample partially observed snapshot \(s_0^{\mathrm{obs}}\!$ and $\Psi(S)$.
    \STATE Build \(\hat{S}_0\!=\!(s_0^{\mathrm{obs}},\,G_\phi(s_0^{\mathrm{obs}},\!\,\Psi(S)))\).
    \STATE Initialize \(S_0\!\leftarrow\!\hat{S}_0\); simulate a rollout of length \(H\!\) under \(\pi_\theta\) to obtain trajectory \(\tau\!\).
  \STATE Compute \(r_{\mathrm{micro}}(\tau;\theta)\) and \(r_{\mathrm{macro}}(\tau;\theta)\) (Eqs.~\eqref{eq:r-micro}--\eqref{eq:r-macro}).
    \STATE Update \(\theta\) to maximize \(J(\theta)\!\) (Eqs.~\eqref{eq:policy-objective}).
\ENDWHILE
\STATE \textbf{Optional:} Repeat Stage~I and Stage~II in a few outer iterations until performance saturates.
\STATE \textbf{Output:} Trained generator \(G_\phi\) and shared policy \(\pi_\theta\).
\end{algorithmic}
\end{algorithm}

\section{Experiment and Case Study}
\label{subsec:case-study}
To further illustrate our proposed solution, we instantiate the general formulation in a controlled ring-road environment.
Consider a circular single-lane roadway with radius \(R\) and \(N\) vehicles moving in the same direction.
The microscopic state of vehicle at time \(t\) is $s_t^i = (\theta_t^i, v_t^i)$
where \(\theta_t^i \in [0,2\pi)\) is the angular position and \(v_t^i\in [10.5,14]m/s\) is the velocity.

From fully observed ring-road trajectories, we instantiate the macroscopic descriptor as
$\Psi(S) = [\,\bar v_{\mathrm{GT}},\, \bar d,\, d_{\min},\, d_{\max},\, v_{\min},\, v_{\max}\,]$,
where $\bar v_{\mathrm{GT}}$ is the fleet-average speed, $\bar d$ is mean spacing between two close vehicles, and
$(d_{\min}, d_{\max})$ and $(v_{\min}, v_{\max})$ are admissible spacing and speed bounds.
These quantities form the macroscopic targets used in the generator loss
and in the macroscopic consistency term of the policy objective.

For this case study, each controllable vehicle uses the local observation
\(o_t^i
=
[v_t^i,\; v_{\mathrm{limit}}(\theta_t^i),\; d_{\mathrm{pre},t}^i,\; \Delta v_t^i]\),
where \(v_{\mathrm{limit}}(\theta_t^i)\) is the local speed limit along the ring,
\(d_{\mathrm{pre},t}^i\) is the headway to the preceding vehicle in angular order,
and \(\Delta v_t^i\) is the relative speed to that preceding vehicle.

The control input coincides with the generic action \(u_t^i\) and is given by \(u_t^i \in [a_{\min}, a_{\max}]\)
with fixed longitudinal acceleration bounds \(a_{\min} = -1.1\) and \(a_{\max} = 0.5\).

Under this ring-road setup and given partially observed scenes \(s_t^{\mathrm{obs}}\) and macroscopic targets $\Psi(S)$, the general objective in Section~\ref{sec:prob} becomes: (i) to learn a generator \(G_\phi\) that reconstructs plausible full states \(\hat{S}_t\),
and (ii) to learn a shared policy \(\pi_\theta\) whose closed-loop behavior is microscopically safe and macroscopically consistent.

\subsection{Data Collection}
\label{subsec:data-collection}

To obtain fully observed training data for this case study,
we simulate traffic on the ring using the Intelligent Driver Model (IDM)~\cite{treiber2000congested}
as the underlying car-following rule.

For each simulation run, \(N\) vehicles are initialized with randomized angular
positions and speeds away from steady state.
IDM parameters are independently sampled from prescribed ranges
to induce heterogeneous behaviors.
Across runs, piecewise-constant speed limits \(v_{\mathrm{limit}}(\theta)\)
are randomized along the ring to create spatially varying operating conditions.
This prevents convergence to a trivial homogeneous equilibrium and yields
diverse, realistic car-following and spacing patterns.

\subsection{Generator Architecture and Autoregressive Rollout}
\label{subsec:gen-arch}

In the ring-road case study, we instantiate the generator \(G_\phi\) with a lightweight autoregressive model
that operates on the microscopic state
\(S_t = \{(\theta_t^i, v_t^i)\}_{i=1}^N\). To be more specific, given a partially observed snapshot \(s_0^{\mathrm{obs}}\) and macroscopic descriptor \(\Psi(S)\), the generator is trained to produce a completed state \(\hat{S}_0=(s_0^{\mathrm{obs}},\hat{s}_0^{\mathrm{hid}})\) via \(\hat{s}_0^{\mathrm{hid}}=G_\phi(s_0^{\mathrm{obs}},\Psi(S))\).

To train the generator, we instantiate the discrepancy as
\(d\!\big(\Psi(\hat{S}_0),\Psi(S)\big)\!=\!\lambda_v L_{\mathrm{speed}}+\lambda_d L_{\mathrm{dist}}\). Thus, the generator is trained with
\begin{equation}
\mathcal{L}_{\mathrm{gen}}(\phi)\!=\!\lambda_v L_{\mathrm{speed}}+\lambda_d L_{\mathrm{dist}}\label{eq:case-loss}
\end{equation}
where \(\lambda_v,\lambda_d\) are scalar weights reported in the case study.
Then we introduced the specific loss term.
The fleet-average speed term is
\(L_{\mathrm{speed}}\!=\!\big(\tfrac{\frac{1}{N}\sum_{i=1}^{N} v_i}{\bar{v}_{\mathrm{GT}}}-1\big)^2\).
When policy generated hidden states are available, we add
\(L_{\mathrm{rec}}\!=\!\tfrac{1}{K}\sum_{j\in\mathcal{I}_{\mathrm{hid}}}\|\hat{s}_0^{\,j}-s_0^{\,j}\|_2^{2}\).Spacing is enforced by a composite penalty\(L_{\mathrm{dist}}\!=\!L_{\mathrm{mean}}+L_{\mathrm{min}}+L_{\mathrm{max}}+L_{\mathrm{var}}\).
Here \(\mathcal{A}\) indexes gaps influenced by generated vehicles.
The mean-gap term is
\(L_{\mathrm{mean}}\!=\!\big(\tfrac{\frac{1}{|\mathcal{A}|}\sum_{i\in\mathcal{A}} d_i}{\bar d}-1\big)^2\).
Lower-bound violations use
\(L_{\mathrm{min}}\!=\!\tfrac{1}{|\mathcal{A}|}\sum_{i\in\mathcal{A}}\big(\tfrac{\max(0,\,d_{\min}-d_i)}{d_{\min}}\big)^2\).
Upper-bound violations use
\(L_{\mathrm{max}}\!=\!\tfrac{1}{|\mathcal{A}|}\sum_{i\in\mathcal{A}}\big(\tfrac{\max(0,\,d_i-d_{\max})}{d_{\max}}\big)^2\).
Dispersion is controlled by
\(L_{\mathrm{var}}\!=\!\operatorname{std}(\{d_i\}_{i\in\mathcal{A}})/\bar d\). Together, these terms align the generator with ideal spacing and speed. Generator training for ring-road case study is summarized in Algorithm~\ref{alg:case-gen-train}.
\begin{algorithm}[H]
  \caption{Generator training in the ring-road case study}
  \label{alg:case-gen-train}
  {\small
  \begin{algorithmic}[1]
    \REQUIRE Training pairs $(s_0^{\mathrm{obs}}, S_0)$ on the ring road; vehicle index set $\mathcal{I}$; descriptor $\Psi(S) = [\bar v_{\mathrm{GT}}, \bar d, d_{\min}, d_{\max}, v_{\min}, v_{\max}]$; weights $\lambda_v, \lambda_d$.
    \STATE Initialize generator parameters $\phi$.
    \FOR{each training iteration}
      \STATE Sample one initial observed state from dataset $s_0^{\mathrm{obs}}$.
      \STATE Complete the hidden states $\hat{s}_0^{\mathrm{hid}} = G_\phi(s_0^{\mathrm{obs}}, \Psi(S))$. Form the completed state $\hat{S}_0 = (s_0^{\mathrm{obs}}, \hat{s}_0^{\mathrm{hid}})$.
      \STATE Compute the generator loss $L_{\mathrm{gen}}(\phi)$ (Eqs.~\eqref{eq:case-loss}) .
      \STATE Update $\phi$ by one gradient step on $L_{\mathrm{gen}}(\phi)$.
    \ENDFOR
  \end{algorithmic}
  }
\end{algorithm}

During generator inference, given a partially observed frame with \(K\) hidden vehicles, we complete the scene via an autoregressive rollout with hard-constraint rejection. At step \(s=1,\dots,K\), the generator proposes \((\hat{\theta}^{(s)},\hat{v}^{(s)})\) given the observed state. We accept the candidate only if it satisfies the distance and velocity bounds \([d_{\min},d_{\max}]\) and \([v_{\min},v_{\max}]\). Otherwise, we resample up to \(T_{\max}\) trials. If none is feasible, we project the last candidate to the nearest feasible bounds. This yields collision-free, macro-consistent completions while keeping observed vehicles unchanged. Generator inference for ring-road case study is summarized in Algorithm~\ref{alg:case-gen-inference}.
\begin{algorithm}[H]
  \caption{Generator inference in the ring-road case study}
  \label{alg:case-gen-inference}
  {\small
  \begin{algorithmic}[1]
    \REQUIRE Trained $G_{\phi}$; partially observed snapshot $s_0^{\mathrm{obs}}$; descriptor $\Psi(S)$; bounds $[d_{\min}, d_{\max}]$, $[v_{\min}, v_{\max}]$; max trials $T_{\max}$; hidden count $K$.
    \STATE Initialize $\hat{S}_0 \gets s_0^{\mathrm{obs}}$.
    \FOR{$s = 1$ to $K$}
      \STATE accepted $\gets$ false.
      \FOR{$t = 1$ to $T_{\max}$}
        \STATE Propose a candidate $(\hat{\theta}^{(s)}, \hat{v}^{(s)}) = G_\phi(\hat{S}_0, \Psi(S))$.
        \STATE Tentatively insert $(\hat{\theta}^{(s)}, \hat{v}^{(s)})$ into $\hat{S}_0$ and compute the affected gaps $d$.
        \IF{all $d \in [d_{\min}, d_{\max}]$ and $\hat{v}^{(s)} \in [v_{\min}, v_{\max}]$}
          \STATE Accept the candidate and update $\hat{S}_0$; set accepted $\gets$ true; \textbf{break}.
        \ENDIF
      \ENDFOR
      \IF{not accepted}
        \STATE \textbf{break} \COMMENT{If no feasible candidate is found after $T_{\max}$ trials}
      \ENDIF
    \ENDFOR
    \STATE Return $\hat{S}_0$.
  \end{algorithmic}
  }
\end{algorithm}

\subsection{Policy Architecture}
\label{subsec:policy-arch}
Episodes are initialized from generator-completed states \(\hat{S}_0\) and rolled out under the shared policy \(\pi_{\theta}\) to produce trajectories \(\tau=\{S_t\}_{t\in\mathcal{T}}\).
We reuse the macroscopic penalties \(L_{\mathrm{speed},t}\) and \(L_{\mathrm{dist},t}\) exactly as defined before, and score an entire policy by aggregating over full rollouts.

We evaluate each rollout with two complementary terms and combine them into a single objective. From the microscopic perspective, on a trajectory \(\tau\) we score agreement with instrumented actions by accumulating their log-likelihood under the policy using \(r_{\mathrm{micro}}(\tau;\theta)=\sum_{i\in\mathcal{I}_{\mathrm{obs}}}\!\sum_{t\in\mathcal{T}}\!\log\!\pi_{\theta}\!\big(u_t^i \mid O(S_t,i)\big)\). From the macroscopic perspective, to encourage aggregate alignment at each step we map the speed/spacing penalties to a bounded reward \(r_{\mathrm{macro}}(\tau;\theta)=\sum_{t\in\mathcal{T}}\! r_{\mathrm{macro},t}\) with \(r_{\mathrm{macro},t}=\big(1+\lambda_v L_{\mathrm{speed},t}+\lambda_d L_{\mathrm{dist},t}\big)^{-1}\) and aggregate over time. The policy maximizes the trajectory-level score \(J(\theta)=\mathbb{E}_{\tau\sim(\pi_{\theta},G_{\phi})}\!\left[r_{\mathrm{micro}}(\tau;\theta)+\eta\,r_{\mathrm{macro}}(\tau;\theta)\right]\), where \(\eta\ge 0\) balances the two terms. We maximize \(J(\theta)\) with Proximal Policy Optimization (PPO)~\cite{schulman2017proximal} using a value-function baseline and generalized advantage estimation. PPO operates on batches of trajectories initialized by \(G_{\phi}\) and uses the clipped surrogate objective. Detailed hyperparameters are provided in Table~\ref{tab:case-hyperparams}.
\begin{table}[h]
\caption{Case study hyperparameters}
\label{tab:case-hyperparams}
\centering
\small
\begin{tabular}{lcl}
\hline
\textbf{Hyperparameter} & \textbf{Symbol} & \textbf{Value} \\
\hline
Ring radius & $R$ & $100~\mathrm{m}$ \\
Vehicle count & $N$ & $5$ \\
Speed range & $[v_{\min},\,v_{\max}]$ & $[10.5,\,14.0]~\mathrm{m/s}$ \\
Target fleet-average speed & $\bar v_{\mathrm{GT}}$ & $12.06~\mathrm{m/s}$ \\
Target mean spacing & $\bar d$ & $126~\mathrm{m}$ \\
Spacing bounds & $[d_{\min},\,d_{\max}]$ & $[115,\,140]~\mathrm{m}$ \\
Acceleration bounds & $[a_{\min},\,a_{\max}]$ & $[-1.1,\,0.5]~\mathrm{m/s^2}$ \\
Hidden vehicles per frame & $K$ & $[0,5]$ \\
Max trials per hidden & $T_{\max}$ & $20$ \\
Speed penalty weight & $\lambda_{v}$ & $0.5$ \\
Spacing penalty weight & $\lambda_{d}$ & $0.5$ \\
Macro reward weight & $\eta$ & $0.3$ \\
Trajectory horizon & $H$ & $10$ \\
\hline
\end{tabular}
\end{table}

\section{Results and discussions}
\label{sec:results}
On the ring-road case study, we demonstrate that our two-stage pipeline learns a shared policy that fuses partial microscopic traces with infrastructure-level aggregates. The resulting policy is aligned with the target behavior both macroscopically and microscopically.
\begin{figure}[!t]
\centering
\includegraphics[trim={0cm 0cm 0cm 0cm}, clip,width=\columnwidth]{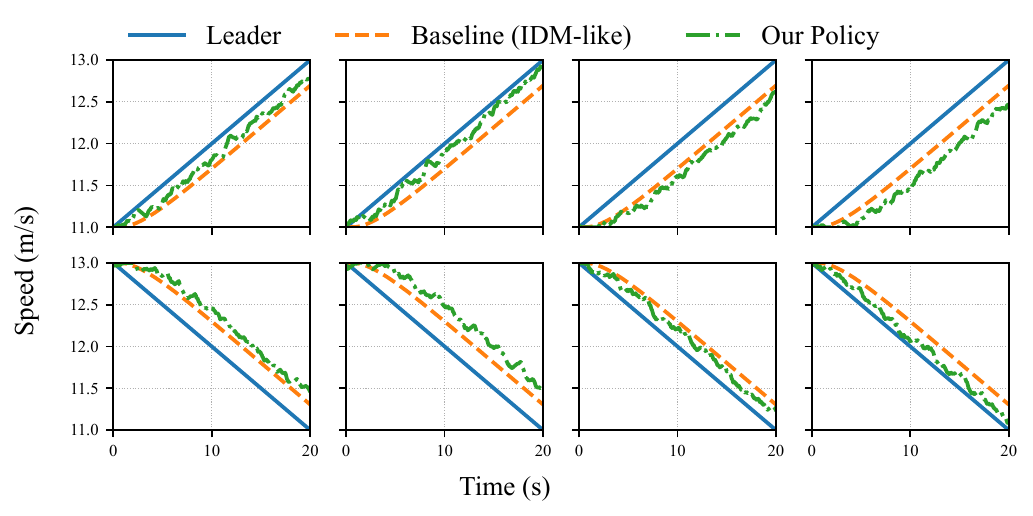}
\vspace{-9mm}
\caption{Leader–follower velocity profiles in two scenarios. Top: leader accelerates 11m/s→13 m/s; bottom: leader decelerates 13m/s→11 m/s. In these plots, the velocity trajectories are drawn in green
(leader), blue (ground-truth IDM follower), and orange (our learned policy follower), respectively.
Across runs, the policy’s average profile follows the ground-truth trend.}
\label{fig:accel_decel_grid}
\vspace{-5mm}
\end{figure}

\subsubsection{Macroscopic Alignment}

We first evaluate whether generator-completed scenes and the learned policy preserve the aggregate traffic statistics defined in Section~\ref{subsec:case-study}.

We analysis the mean and standard deviation of longitudinal speed and distance to the preceding vehicle using the same simulation settings. For simulation rollouts, we choose the number of unobserved vehicle to be \(K=1\)--4 per scenario.

For unobserved counts \(K \in \{1,2,3,4\}\), the learned policy maintains a mean speed close to the ground-truth mean of \(12.06\,\mathrm{m/s}\), yielding \(11.55\), \(11.54\), \(11.52\), and \(11.53\,\mathrm{m/s}\), respectively; the largest absolute deviation is \(0.54\,\mathrm{m/s}\). The speed standard deviation increases from \(0.37\,\mathrm{m/s}\) (ground-truth) to \(0.78\)--\(0.80\,\mathrm{m/s}\). The mean spacing remains \(125.66\,\mathrm{m}\) for all \(K\). The spacing standard deviation changes only slightly, from \(34.60\,\mathrm{m}\) (ground-truth) to at most \(34.79\,\mathrm{m}\), an increase of \(0.19\,\mathrm{m}\).

Across all values of \(K\), the mean speeds remain close to the ground-truth value and the spacing statistics stay essentially unchanged, while the increases in variance are mild. This shows that the generator reconstructs hidden vehicles without distorting macroscopic density, and that the learned policy, operating on these completions, maintains realistic macroscopic behavior under partial observations.

\subsubsection{Microscopic Behavior of the Learned Policy}

We next examine whether the learned policy exhibits meaningful microscopic responses consistent with the macroscopic alignment.
To analyze its driving behavior, we construct two leader–follower scenarios in which the leader either accelerates or decelerates according to a prescribed
velocity profile. In each scenario, the lead vehicle follows this fixed profile, while the
following vehicle is controlled by either the ground-truth IDM-based policy or our learned policy.
The follower starts 126 m behind the leader and shares the same initial speed.

Fig.~\ref{fig:accel_decel_grid} visualizes representative leader-follower trials under leader acceleration-only and leader deceleration-only scenarios.
The learned policy does not exactly replicate the IDM trajectory, but its responses are coherent: it accelerates and decelerates in the appropriate phases, maintains safe headways, and avoids unrealistic gap collapse or divergence.

Over multiple trials, some sampled trajectories lie slightly above the ground-truth velocity profile and some lie slightly below, in both acceleration-only and deceleration-only scenarios.
This variability is structured and physically plausible, rather than noisy.
As a result, when aggregating over trials, the ensemble of rollouts matches the target macroscopic statistics in both mean and dispersion.
In other words, the policy is macroscopically aligned precisely because its microscopic behaviors form a balanced family of realistic car-following patterns, rather than collapsing to a single deterministic trace.

Together, the macroscopic statistics and shared policy analysis demonstrate that the proposed generator-policy pipeline is macroscopically consistent and microscopically meaningful. It preserves aggregate traffic statistics while producing plausible local interactions that explain those statistics.

\section{Conclusion and Future Work}
In this paper, we have presented a macro-micro framework for learning a shared driving policy from heterogeneous traffic data, combining partially observed microscopic trajectories with macroscopic statistics. The approach uses a generator to complete hidden vehicle states and a shared policy trained on these completed scenes to achieve consistency with both individual behaviors and aggregate traffic patterns. We have demonstrated the approach in a simple controlled ring-road case study. Future work will focus on extending this framework to multi-lane roads, intersections, and network-scale settings with richer heterogeneity. Methods will be further evaluated using real-world datasets that provide both infrastructure-level measurements and subset-based microscopic trajectories within the same environment.

\bibliographystyle{IEEEtran}
\bibliography{ref}

\vspace{12pt}

\end{document}